\documentclass{JHEP3}
\usepackage{amsmath}\usepackage{amssymb}
\usepackage{cancel}

\newcommand{\bea}{\begin{eqnarray}}
\newcommand{\eea}{\end{eqnarray}}
\newcommand{\bi} {\begin{itemize}}
\newcommand{\ei} {\end{itemize}}

\def\half{{\frac{1}{2}}}
\def\nl{\nonumber \\}
\def\br{\overline}
\def\dalpha{{\dot\alpha}}
\def\dbeta{{\dot\beta}}

\def\AA { {A}}
\def\AAi {({A}^{-1})}
\def\BB {{\hat{B}}}
\def\BBi {(B^{-1})}
\def\LL {\mathcal {L}}
\def\VV {v}
\def\l{\left(}
\def\r{\right)}
\def\ee{\varepsilon}
\def\AVD{\hat\nabla}
\def\bra{\left\langle}
\def\ket{\right\rangle}
\def\vev#1{\bra #1 \ket}
\def\cT{{\hat T}}
\def\cD{{\cal D}}

\title{Supersymmetric sound in fluids}

\author{
Carlos Hoyos,${}^1$
Boaz Keren-Zur${}^2$
  and Yaron Oz${}^1$\\
  ${}^1$Raymond and Beverly Sackler Faculty of Exact Sciences \\
School of Physics and Astronomy \\
Tel-Aviv University, Ramat-Aviv 69978, Israel\\
${}^2$Institut de Th\'eorie des Ph\'enom\`enes Physiques, EPFL,\\
CH-1015 Lausanne, Switzerland\\
E-mails: \email{choyos@post.tau.ac.il}, \email{boaz.kerenzur@epfl.ch},
  \email{yaronoz@post.tau.ac.il}
}

\abstract{We consider the hydrodynamics of supersymmetric fluids. Supersymmetry is  broken spontaneously
and the low energy spectrum includes a fermionic massless mode, the $\mathit{phonino}$.
We use two complementary approaches to describe the system: First, we construct a generating functional from which we derive the equations of motion of the fluid and of the phonino propagating through the fluid. We write the form of the leading corrections in the derivative expansion, and show that the so called diffusion terms  in the supercurrent are in fact not dissipative. 
Second, we use an effective field theory approach which utilizes a non-linear realization of supersymmetry to analyze the interactions between phoninos and phonons, and demonstrate the conservation of entropy in ideal fluids.
We comment on possible phenomenological consequences for gravitino physics in the early universe.
}


\begin{document}

\section{Introduction}

Spontaneously broken symmetries play an important role in the physics of fluids.
Phonons can be considered as the Goldstone modes of Lorentz symmetry, which is broken by the choice of the fluid's rest frame. Another example is the phenomenon of superfluidity, where the spontaneous breaking of a $U(1)$ global symmetry leads to the appearance of a non-dissipative flow in the fluid.

Supersymmetry is broken at finite temperature since the thermal ensemble differentiates between the statistics of fermions and bosons. Interestingly, this symmetry is not restored at high temperature \cite{Das:1978rx}.
It is well known that when supersymmetry is spontaneously broken, the low energy spectrum contains a fermionic massless mode, known at zero temperature as the $\mathit{goldstino}$.
A similar situation happens in supersymmetric fluids at finite temperature, where a
long range fermionic fluctuation appears.
This mode is called $\mathit{phonino}$.
The existence of this mode was demonstrated by using the real-time
formalism in~\cite{Boyanovsky:1983tu}. Its dispersion relation was calculated for strongly coupled supersymmetric plasma using the AdS/CFT correspondence in  \cite{Policastro:2008cx,Gauntlett:2011wm,Kontoudi:2012mu}.
At leading order in momenta, the phonino satisfies a linear dispersion relation with velocity (see also~\cite{Leigh:1995jw,Kratzert:2003cr}):
\bea
\label{eq_phonino}
v_G=\left | \frac{P}{\varepsilon}\right| \ ,
\eea
where $P$ is the pressure and $\varepsilon$ is the energy density.
In simple cases where $P\propto \varepsilon$, the speed of the phonino is the square of the speed of (first) sound.
At zero temperature $|P| = |\varepsilon|$ and the phonino is the goldstino, a free relativistic fermion.

The existence of this mode can be understood as a consequence of a supersymmetric Ward-Takahashi identity for the supercurrent two-point function \cite{wtid}\footnote{In our conventions the supercurrent has a relative factor $1/\sqrt{2}$ respect to the one originally used. This removes a factor of $2$ from the right hand side of the WT identity.}:
\begin{equation}\label{WTsusy}
\partial_\mu\vev{T\{S^\mu(x) \bar{S}^\nu(y)\}}=\delta^{(4)}(x-y)\,\vev{T^{\nu\rho}}\sigma_\rho  \ .
\end{equation}
Going to momentum space and assuming a constant energy-momentum tensor the identity becomes
\begin{equation}\label{WTSS}
-i k_\mu \Gamma^{\mu\nu}_{SS}=\vev{T^{\nu\rho}}\sigma_\rho \ .
\end{equation}
In order to satisfy this identity for all momenta we need that $\Gamma_{SS} \sim 1/k$, which is possible if there is a massless fermionic mode, the goldstino/phonino. Let us assume that the system is at thermal equilibrium and that the energy-momentum tensor takes the form of that of an ideal fluid
\begin{equation}
\vev{T^{\mu\nu}}={\rm diag}\,(\varepsilon,P,P,P) \ .
\end{equation}

As in the case of spontaneous breaking of supersymmetry, we expand the supercurrent $S^{\mu}$ ($\bar{S}^{\mu}$) at low energies in derivatives of the phonino field $\bar G$ ($G$) :
\begin{equation}
S^0=iF_t \sigma^0 \bar G+\cdots, \ \ S^i=iF_s \sigma^i \bar G+\cdots \ ,
\end{equation}
where we have assumed rotational invariance but not Lorentz invariance. In these conventions $G$ is a Weyl fermion. The identity \eqref{WTSS} becomes
\begin{align}
&-i (F_t k_0 \sigma^0+F_s k_i\sigma^i) \vev{G\bar{G}} F_t \sigma^0 =\varepsilon \sigma^0,\\
&-i (F_t k_0 \sigma^0+F_s k_i\sigma^i) \vev{G\bar{G}} F_s \sigma^i =-P \sigma^i \ ,
\end{align}
The equations require that the propagator of the phonino has the form
\begin{equation}
\vev{G\bar{G}}=\frac{1}{F_t}\frac{i\left( k_0 \sigma^0+ v_G k_i\sigma^i\right)}{k_0^2-v_G^2 k_i^2} \ ,
\end{equation}
where
\begin{equation}
F_t=\varepsilon, \ F_s=-P, \ v_G=\frac{F_s}{F_t}=-\frac{P}{\varepsilon} \ .
\end{equation}
There is some arbitrariness in the normalization, this choice corresponds to the definition of the supercurrent

\begin{equation}
S^\mu=iT^\mu_\nu\sigma^\nu \bar G+\cdots
\end{equation}
such that the conservation of the supercurrent becomes
\begin{equation}
0=\partial_\mu S^\mu=iT^\mu_\nu\sigma^\nu\partial_\mu \bar G+\cdots
\end{equation}

At zero temperature $T^\mu_\nu=-\varepsilon\delta^\mu_\nu$ and the propagator of the phonino becomes that of the goldstino. In this case the energy density is determined by the scale of supersymmetry breaking $\varepsilon=|f|^2$. When Lorentz invariance is broken, $|v_G|\neq 1$ is the velocity of the phonino.
Note also that the conformal Ward identity for the supercurrent agrees with the tracelessness of the energy-momentum tensor in a CFT
\begin{equation}
\sigma_\mu S^\mu\propto (T^\mu_\mu)\bar G+\cdots=0 \ .
\end{equation}
These two equations are valid to lowest order at the ideal level even out of equilibrium, although there can be derivative corrections beyond the ideal level.

In this work we will be studying the hydrodynamic regime of supersymmetric field theories, that is the
the combined effective theory of the phonino and the fluid. In particular we will be interested in the motion of the phonino through the fluid itself and in its interactions with other long range modes in the fluid, the phonons.
In order to describe the motion of the phonino through the fluid, we will begin by constructing the effective action following a method inspired by the recent works \cite{Jensen:2012jh,Banerjee:2012iz}\footnote{See also \cite{Loganayagam:2012pz}, where the partition function is constructed from a microscopic theory with free fermions.} to derive relations between transport coefficients from a generating functional depending on external metric and gauge fields. We will adapt the method to include gravitino sources and the phonino.
From the resulting generating functional one can derive the equations of motion of the fluid and of the phonino propagating through the fluid. Since the gravitino sources are included, it can also be directly interpreted as the contribution to the effective action of the gravitino from the fluid. As we will see, a mass for the gravitino appears at the order of the first derivative corrections of the fluid.

We note that supersymmetric hydrodynamics has been studied in
\cite{Kovtun:2003vj}. Our analysis is rather different. In \cite{Kovtun:2003vj} the
concepts of "classical" supersymmetric charge and chemical potential are introduced and are used to define the constitutive relations with which one can express the supercurrent. This is done is analogy to the hydrodynamic description of bosonic currents such as the stress-energy tensor
and global symmetry currents. However, one cannot define consistently classical fermionic charges as the expectation values
of fermionic operators is zero.
We therefore propose that the correct analysis of the hydrodynamics of supersymmetric field theories is not that of a flow of a fermionic charge, but rather the dynamics
of the phonino in the normal fluid.

Although in principle one can derive all the relevant information from this generating functional, in order to study interactions of the phonino with the phonons in the fluid it is better to use a slightly different set of variables, that in particular will allow us to make contact with the effective actions at zero temperature of the goldstino existing in the literature. The approach is based on the Akulov-Volkov formalism for non-linearly realized supersymmetry (NLRS) \cite{Volkov:1972jx}.
The main results of this analysis 
are the following: first, we introduce a formulation to relate NLRS to hydrodynamics via conserved currents which is shown to be consistent on both ends - it is a valid hydrodynamics description of ideal fluids in the sense that
entropy is shown to be conserved, and it has the correct transformation properties under supersymmetry.
The important new constraint we impose is that both the currents and their conservation equations transform in a way which is consistent with the SUSY algebra. This leads to a unique construction of the physical currents in terms of currents dressed with appropriate goldstino factors. Since this construction is dictated by symmetry, 
it is valid even in the absence of a Lagrangian description for the microscopic system. Secondly, we use this formalism to study the dynamics of phoninos in ideal fluids. We find the phonino dispersion relations and compute the expressions for the leading interactions between phoninos and phonons.

The paper is divided into two sections -- one for each of the two formalisms mentioned above.
We conclude with a short discussion of the phenomenological implications of this analysis.
A review of non-linearly realized supersymmetry and details of some of the computations appearing in the paper are given in the appendix.

\section{Generating functional with a phonino}

We will consider a relativistic fluid in the presence of external sources. We will assume that all the microscopic degrees of freedom have been integrated out and that we can use a hydrodynamic approximation to describe the properties of the system. If the underlying theory is supersymmetric we expect that a collective massless fermionic fluctuation exists, the phonino. The motion of the phonino should be included in the hydrodynamic description, and as a first approximation we can treat the phonino as propagating through a thermal medium but not affecting its properties. The effective action of the phonino will depend on the properties of the medium and on external sources, and we can use supersymmetry to constrain its form. The partition function of the theory will take the form
\begin{equation}
{\cal Z}=\int DG D\bar{G} e^{iW[G]},
\end{equation}
where $G$ is the phonino and $W$ is determined by the temperature and possible external sources. In principle $W$ is a non-local functional, but it has a local expansion when static configurations (in the rest frame) are considered as explained in \cite{Jensen:2012jh,Banerjee:2012iz}.

In order to describe the effect of the medium, the temperature and the chemical potential are treated as external sources and $W$ becomes the generating functional for them. In the absence of chemical potential, the generating functional takes the form
\begin{equation}
W=\int d^d x \sqrt{-g} P(T)+\cdots
\end{equation}
Where $T=T_0/\sqrt{-V^2}=T_0/\sqrt{-g_{\mu\nu}V^\mu V^\nu}$ is the temperature and $V^\mu$ is proportional to the velocity of the fluid.

The energy-momentum tensor is
\begin{equation}
T^{\mu\nu}=\frac{2}{\sqrt{-g}}\frac{\delta W}{\delta g_{\mu\nu}}=Pg^{\mu\nu}+T\frac{\partial P}{\partial T}\frac{V^\mu}{\sqrt{-V^2}}\frac{V^\nu}{\sqrt{-V^2}}=Pg^{\mu\nu}+Ts\, u^\mu u^\nu.
\end{equation}
The equations of motion can be deduced from diffeomorphism invariance of the underlying theory. Although the background metric breaks physical diffeomorphism invariance, the partition function is invariant up to anomalies under a combined transformation of the dynamical fields and the external sources. This implies that the generating functional should be invariant up to anomalies under transformations of the sources
\begin{equation}
\delta_a g_{\mu\nu}=\nabla_\mu a_\nu+\nabla_\nu a_\mu.
\end{equation}
Then,
\begin{equation}
0=\delta_a W=\int d^d x \frac{\delta W}{\delta g_{\mu\nu}}\delta_a g_{\mu\nu}\ \ \Rightarrow \ \ \nabla_\mu T^{\mu\nu}=0.
\end{equation}

Suppose now that there is a global symmetry with a chemical potential\footnote{This expression is not explicitly gauge-invariant. A gauge-invariant expression involves a Wilson line along the time direction defined by the Killing vector $V^\mu\partial_\mu$. Fixing the gauge so that only time-independent transformations are allowed, the chemical potential takes the form \eqref{cp}.}
\begin{equation}\label{cp}
\mu=\frac{V^\mu A_\mu}{\sqrt{-V^2}}.
\end{equation}
The current is defined as
\begin{equation}
J^\mu=\frac{1}{\sqrt{-g}}\frac{\delta W}{\delta A_\mu}=\frac{\partial P}{\partial \mu} \frac{V^\mu}{\sqrt{-V^2}}=\rho u^\mu.
\end{equation}
invariance of the generating functional under gauge transformations $\delta_\lambda A_\mu=\partial_\mu\lambda$ imposes the following condition
\begin{equation}
0=\delta_\lambda W =-\int d^d x  \sqrt{-g} \lambda\nabla_\mu J^\mu.
\end{equation}
If the global symmetry is spontaneously broken there is a Goldstone boson $\phi$. Under a global symmetry transformation $\delta_\lambda\phi=\lambda$. The chemical potential can be defined as
\begin{equation}
\mu= \frac{V^\mu(A_\mu-\nabla_\mu\phi)}{\sqrt{-V^2}},
\end{equation}
which is explicitly gauge invariant and gives the Josephson condition $u^\mu\nabla_\mu\phi=-\mu$ in the absence of sources.
A new scalar quantity on which the generating functional can depend is $(A_\mu-\nabla_\mu\phi)^2$.

If the underlying theory is supersymmetric, we can extend the same arguments to supersymmetric transformations. The difference is that now we should consider more general external sources, including a gravitino. The generating functional is a function of the vierbein, spin connection and gravitino background values (we do not consider global bosonic symmetries for the moment)
\begin{equation}
W[e_\mu^a,\omega_\mu^{\,ab},\Psi_\mu^\alpha,\bar{\Psi}_\mu^{\dot \alpha}]=\int d^dx \sqrt{-g}P(T,\mu^\alpha,\bar{\mu}^{\dot{\alpha}}),
\end{equation}
Here $\alpha$ refers to spinor indices and we denote
\begin{equation}
\mu^\alpha=\frac{V^\mu}{\sqrt{-V^2}}\Psi_\mu^\alpha \ .
\end{equation}
The generating functional should be understood as having an expansion in the Grassmann fields. $\Psi_\mu$ to be a Weyl spinor in the notation of \cite{Wess:1992cp}.

Under a supersymmetric transformation, the variation of the sources is (suppressing spinor indices)
\begin{align}\label{susytr}
& \delta_\xi e_\mu^a =i(\Psi_\mu\sigma^a\bar{\xi}-\xi\sigma^a\bar{\Psi}_\mu),\\
& \delta_\xi \omega_\mu^{\,ab}=0,\\
& \delta_\xi \Psi_\mu=-2{\cal D}_\mu\xi,\\
& \delta_\xi \bar{\Psi}_\mu=-2{\cal D}_\mu\bar{\xi}.
\end{align}
The derivative ${\cal D}_\mu$ includes the spin connection.
These transformations are actually special for Poincar\'e symmetry. Had we considered for instance a theory in anti-de Sitter space the transformations would be different. Introducing external gauge fields will also modify the supersymmetric transformations.

A general supersymmetric transformations of the generating functional leads to
\begin{align}\label{susyvar0}
&\delta_\xi W=\int d^4 x\left[\frac{\delta W}{\delta e_\mu^a}\delta e_\mu^a+ \frac{\delta W}{\delta \bar{\Psi}_\mu}\delta\bar{\Psi}_\mu+\delta \Psi_\mu\frac{\delta W}{\delta \Psi_\mu}\right].
\end{align}
\begin{align}\label{susyvar}
&0=\delta_\xi W=\int d^4x\,\sqrt{-g}\left[iT^{\mu\nu}\Psi_{\mu}\sigma_{\nu}\bar{\xi}+2\xi {\cal D}_\mu S^\mu  +h.c.\right] \ .
\end{align}
Therefore, the conservation equations for the supercurrent are: 
\begin{equation}
2 {\cal D}_\mu S^\mu =i T^{\mu\nu}\sigma_{\mu}\bar{\Psi}_{\nu}, \ \ 2 {\cal D}_\mu \bar{S}^\mu =-i T^{\mu\nu}\Psi_{\mu}\sigma_{\nu} \ .
\end{equation}

Now we can take into account the effective spontaneous breaking of supersymmetry, as it happens at finite temperature, by introducing the phonino field, whose transformation under supersymmetry is
\begin{equation}\label{transfphonino}
\delta_\xi G = \xi, \ \ \delta_\xi\bar{G}=\bar{\xi}.
\end{equation}
The scale of supersymmetry breaking appears in the definition of the supercurrent. This field is analogous to the phase of the condensate in spontaneous breaking of global symmetries.

The effective action of the phonino will be determined by the original generating functional in an analogous way as some terms in the effective action of Goldstone bosons are determined by the generating functional for gauge fields. With the phonino we can construct the SUSY-invariant combinations
\begin{equation}
\psi_\mu\equiv \Psi_\mu+2{\cal D}_\mu G, \ \ \bar{\psi}_\mu=\bar{\Psi}_\mu+2\cD_\mu \bar{G},
\end{equation}
for the gravitino sources. For the vierbein we can form the combination
\begin{equation}
E_\mu^a=e_\mu^a-i(\Psi_\mu\sigma^a\bar{G}-G\sigma^a\bar{\Psi}_\mu)-2i(\cD_\mu G\sigma^a\bar{G}-G\sigma^a\cD_\mu\bar{G}).
\end{equation}
The transformation of the dressed vierbein is
\begin{equation}
\delta_\xi E_\mu^a=-2i(\cD_\mu G \sigma^a \bar{\xi}-\bar{\xi}\sigma^a\cD_\mu\bar{G}).
\end{equation}
We can also write it as
\begin{equation}
\label{eq_vierbein}
E_\mu^a=e_\mu^a-i(\psi_\mu\sigma^a\bar{G}-G\sigma^a\bar{\psi}_\mu)  \ .
\end{equation}
This choice for the dressing of the vierbein leads to an action whose supersymmetry variation is independent of the gravitino and, as shown below, gives the correct equations of motion.
This approach is similar to the AV formalism we will use in the next sections, but the hydrodynamic variables will be described differently.
Note also, that in the current approach the SUSY transformations are linear and the variation of the
generating functional gives the conservations laws, while  in the AV formalism  the SUSY transformations are nonlinear under which the effective action is invariant.

The dependence on the metric is changed to a dependence on the covariant combination $g_{\mu\nu}\to G_{\mu\nu}=\eta_{ab} E_\mu^a E_\nu^b$. Then, the pressure depends on the dressed temperature
\begin{equation}
\cT= \frac{T_0}{\sqrt{-G_{\mu\nu}V^\mu V^\nu}},
\end{equation}
where
\begin{equation}
G_{\mu\nu}V^\mu V^\nu=g_{\mu\nu}V^\mu V^\nu\left(1-2i u^a(\mu \sigma_a \bar{G}-G\sigma_a \bar{\mu})-\eta_{ab}(\mu \sigma^a \bar{G}-G\sigma^a \bar{\mu}) (\mu \sigma^b \bar{G}-G\sigma^b \bar{\mu})\right).
\end{equation}
The velocity and the fermionic chemical potential are defined in the usual way, as we will also do with the temperature
\begin{equation}
u^\mu=\frac{V^\mu}{\sqrt{-V^2}}, \ \ \mu=u^\mu\psi_\mu, \ \ T=\frac{T_0}{\sqrt{-V^2}}.
\end{equation}
where we have defined $V^2=g_{\mu\nu}V^\mu V^\nu$.

In the absence of sources the definition of the chemical potential gives a relation $u^\mu\psi_\mu=u^\mu {\cal D}_\mu G=\mu $ similar to the one found in a superfluid. In the generating functional the covariant form of the chemical potential should appear
\begin{equation}
\label{eq_hat_mu}
\hat \mu=\frac{V^\mu}{\sqrt{-G_{\mu\nu}V^\mu V^\nu}}\psi_\mu \ .
\end{equation}

The pressure is a function of the temperature and fermionic bilinears $P(\cT,\psi^2,\hat \mu,\bar{{\hat \mu}})$, but not all the terms are of the same order. We are doing an expansion in small derivatives $\nabla_\mu \sim \epsilon$. In order to keep all contributions to the vierbein of the same order $E_\mu^a\sim 1$ and assuming that the velocities are of the same order $V^\mu\sim 1$, we should fix the scaling of the phonino to $G\sim 1/\sqrt{\epsilon}$. In this case $\psi_\mu \sim \sqrt{\epsilon}$, so bilinear contributions to the stress-energy tensor are suppressed $\sim \epsilon$. Similarly, this scaling behavior ensures that contributions to the supercurrent which are linear in $\psi$ are suppressed by a factor of $\epsilon$ with respect to the leading contribution, which is proportional to $G$.
We conclude that the contributions related to $\psi$ and $\hat \mu$ are of order $\epsilon$, the same as the contributions of bosonic fields that contain one derivative.

We now  define the generating functional to leading order as
\begin{equation}
W=\int d^4 x E P(\cT),
\end{equation}
where $E$ is the veilbein determinant.
Under a supersymmetric variation
\begin{equation}
\delta_\xi W = \int d^4 x\,E\left(2i T^\mu_{\ a}\cD_\mu G \sigma^a \bar \xi+h.c.\right)~,
\end{equation}
where the energy-momentum tensor has been defined as
\begin{equation}
T^\mu_a =\frac{1}{E}\frac{\delta W}{\delta E_\mu^a}.
\end{equation}
Therefore, supersymmetry requires that the following condition is satisfied
\begin{equation}\label{susycond}
T^\mu_{\ a}\cD_\mu G \sigma^a=0~.
\end{equation}

Using the condition \eqref{susycond} and the Fierz identities
\begin{equation}
\chi\sigma^a\bar{\psi} = -\psi\sigma^a\bar{\chi}, \ 
\end{equation}
the generating functional satisfies the Wess-Zumino consistency condition
\begin{equation}
(\delta_\eta\delta_\xi-\delta_\xi\delta_\eta)W=2i\int d^4 x\,ET^\mu_{\ a}\cD_\mu(\eta\sigma^a\bar{\xi}-\xi\sigma^a\bar{\eta}).
\end{equation}

Since $E_\mu^a$ transforms under diffeomorphisms in the same way as $e_\mu^a$, diffeomorphism invariance of the action requires that $T^\mu_a$ is conserved. The conservation equation is, in flat space
\begin{equation}
\partial_\mu\left(E T^\mu_{\ a}\right)=0 \ .
\end{equation}
Although there is a factor depending on the vierbein, its derivative is zero in the absence of gravitino sources
\begin{equation}
E E_b^\nu T^\mu_{\ a}\partial_\mu E_\nu^b=-4 iE E_b^\nu T^\mu_{\ a}\partial_\mu G\sigma^a\partial_\nu \overline{G}=0 \ .
\end{equation}
Where we have used that
\begin{equation}
T^\mu_{\ a}\partial_\mu G\sigma^a=0 \ .
\end{equation}

We can use the vierbeins to define the energy-momentum purely in the orthogonal frame
\begin{equation}
T^a_{\ b}=E_\mu^a T^\mu_{\ b} \ .
\end{equation}
The explicit form of the energy-momentum tensor is
\begin{equation}
T_{ab}=\eta_{ab} P(\cT)+\cT\frac{\partial P}{\partial \cT} \frac{\eta_{ac}\eta_{bd} E_\mu^c V^\mu E_\nu^d V^\nu}{-\eta_{ab}E_\mu^a V^\mu E_\nu^b V^\nu}.
\end{equation}
We can simplify further this expression using that
\begin{equation}
\label{eq_dressing_V}
E_\mu^a V^\mu = V^a -iV^\mu(\psi_\mu\sigma^a\bar{G}-G\sigma^a\bar{\psi}_\mu)
=V^a +\sqrt{-V^2}\zeta^a
=\sqrt{-V^2}(u^a+\zeta^a),
\end{equation}
where $V^a=e_\mu^a V^\mu$ as usual and we have defined
\begin{equation}
\zeta^a=-i(\mu \sigma^a\bar{G}-G\sigma^a \bar{\mu}) \ .
\end{equation}
Note that the dressed temperature is
\begin{equation}
\cT=\frac{T}{\sqrt{-\eta_{ab}(u^a+\zeta^a)(u^b+\zeta^b)}} \ .
\end{equation}

Assuming there is no external gravitino, there are only four independent components in the Grassmann field $G$, so there can be up to $O(\zeta^2)$ terms.
For the temperature this means
\begin{equation}
\cT= T\left(1+u^a\zeta_a+\frac{1}{2}(\eta^{ab}+3u^a u^b)\zeta_a\zeta_b\right).
\end{equation}
The pressure then can also be expanded as
\begin{equation}
P(\cT)=P(T)+\frac{\partial P}{\partial T}\left(u^a\zeta_a+\frac{1}{2}(\eta^{ab}+3u^a u^b)\zeta_a\zeta_b\right)+\frac{1}{2}\frac{\partial^2 P}{\partial T^2}u^a u^b \zeta_a\zeta_b.
\end{equation}
Finally, the energy momentum tensor takes the form
\begin{align}
\notag & T_{ab}=\eta_{ab}\left(P(T)+a_1 u^c\zeta_c+(a_2 \eta^{cd}+a_3 u^c u^d)\zeta_c\zeta_d \right)\\
\notag &+u_a u_b\left(T\frac{\partial P}{\partial T}+b_1 u^c\zeta_c+(b_2 \eta^{cd}+b_3 u^c u^d)\zeta_c\zeta_d\right)\\
&+(u_a\zeta_b+u_b\zeta_a)\left( T\frac{\partial P}{\partial T}+c_1 u^c\zeta_c\right)+\zeta_a\zeta_bT\frac{\partial P}{\partial T}.
\end{align}
Where
\begin{align}
\notag &a_1=\frac{\partial P}{\partial T},  \ \ \ a_2=\frac{a_1}{2},  \ \ \ a_3=3a_1+\frac{\partial a_1}{\partial T},   \\
\notag &b_1=T\left(2a_1+\frac{\partial a_1}{\partial T}\right),  \ \ \ b_2=\frac{b_1}{2},   \ \ \ b_3 =T\left(4a_1+\frac{7}{2} \frac{\partial a_1}{\partial T}+\frac{\partial^2 a_1}{\partial T^2}\right),  \\
&c_1=b_1.
\end{align}

The first derivative corrections depending on the phonino are easy to compute. First note that the field $\psi_\mu$ is in a $\left(\frac{1}{2},0 \right)\otimes \left(\frac{1}{2},\frac{1}{2}\right)=\left(0,\frac{1}{2} \right)\oplus \left(1,\frac{1}{2} \right)$ representation of the Lorentz group, while $\bar{\psi}_\mu$ is in $\left(0,\frac{1}{2} \right)\otimes \left(\frac{1}{2},\frac{1}{2}\right)=\left(\frac{1}{2},0 \right)\oplus \left(\frac{1}{2},1 \right)$. Therefore, there are four possible scalar bilinear combinations
\begin{align}
G^{\mu\nu}\psi_\mu \psi_\nu , \ \ G^{\mu\nu}\bar{\psi}_\mu \bar{\psi}_\nu, \ \ \psi_\mu \sigma^{[\mu\nu]} \psi_\nu, \ \ \bar{\psi}_\mu \bar{\sigma}^{[\mu\nu]} \bar{\psi}_\nu  ,
\end{align}
all of which are of order $\sim \epsilon $ in our expansion. We can add these terms in the pressure with arbitrary coefficients depending on the dressed temperature
\begin{equation}
P=P_0(\cT)+P_1(\cT)\hat \mu^2 +P_2(\cT)G^{\mu\nu}\psi_\mu \psi_\nu +P_3(\cT) \psi_\mu \sigma^{[\mu\nu]} \psi_\nu+h.c.+\cdots
\label{Ps}
\end{equation}
Note, that the new terms can also be interpreted as bilinear terms for the gravitino field induced by the fluid. In particular, the term proportional to $P_3(\cT)$ has the form of a mass for a Rarita-Schwinger field, while $P_2(\cT)$ and could be interpreted as a mass for the spin-$1/2$ components of the gravitino field. Note that both coefficients can be complex in general, for instance
\begin{equation}
P_2(\cT)G^{\mu\nu}\psi_\mu \psi_\nu+h.c.={\rm Re}\, P_2(\cT)\left(G^{\mu\nu}\psi_\mu \psi_\nu+G^{\mu\nu}\bar{\psi}_\mu \bar{\psi}_\nu\right)+i {\rm Im}\, P_2(\cT)\left(G^{\mu\nu}\psi_\mu \psi_\nu-G^{\mu\nu}\bar{\psi}_\mu \bar{\psi}_\nu\right).
\end{equation}
Both terms can contribute to the dispersion relation of the phonino, adding a contribution quadratic in the momentum which has a real part proportional to  ${\rm Re}\, P_2(\cT)$ and a imaginary part proportional to ${\rm Im}\, P_2(\cT)$. In the conformal case  $P_2(\cT)$ and $P_3(\cT)$ are not independent. $P_2(\cT)$ has been calculated for particular examples in the AdS/CFT framework \cite{Policastro:2008cx,Gauntlett:2011wm,Kontoudi:2012mu} and it modifies the phonino
dispersion relation. It was found that for ${\cal N}=4$ SYM, ${\rm Re}\, P_2(\cT)=0$ and ${\rm Im}\, P_2(\cT)\neq 0$, implying that this is a damped mode. However, it is clear from our construction that although it apppears at the first derivative order it is not a dissipative term in the sense that it does not contribute to the entropy current.

In the following sections we will introduce a slightly different formalism that we will use to compute the interactions between phoninos and phonons. Although the results are equivalent in the two formalisms and there are many parallelisms, the differences are important enough that we will give a detailed description of it.

\section{Effective field theory approach for phonino dynamics}

Phonino dynamics can be studied using an effective field theory. This approach is based  on a Lagrangian description, and therefore valid only for ideal fluids, but it will give a new perspective on the results of the generating functional formalism and enable us to investigate the interactions with the other long range modes in the fluid, the phonons.
As in the previous section, we write the most general action allowed by symmetry, and the effects of spontaneously broken supersymmetry are manifested in the appearance of the fermionic zero mode. The framework we will use to incorporate the phonino into the dynamics of the fluid is the Akulov-Volkov (AV) formalism for non-linearly realized supersymmetry (NLRS), which is reviewed in appendix \ref{app_NLRS}
\cite{Volkov:1972jx}.\footnote{Another possible formalism for NLRS is the framework of constrained superfields which has recently gained renewed attention thanks to the work of \cite{Komargodski:2009rz}.
We chose to avoid the usage of superfields because at nonzero temperature they require non-trivial boundary conditions in superspace.}
At this point we would like to emphasize three points related to this framework. First, we define the standard non-linear realization of supersymmetry for a generic field $\phi$ as follows:
\bea
\label{eq_standard_NLRS}
\delta_\xi\phi&=&-i\l G \sigma^\mu \br \xi-\xi \sigma^\mu \br G\r\partial_\mu \phi~.
\eea
In the appendix it is shown that the transformation properties of the Goldstino are such that this transformation rule satisfies the supersymmetry algebra.

Secondly, we introduce the AV vierbein
 \bea
\label{eq_AA}
\AA^{~a}_{\mu}&\equiv&\delta^{a}_{\mu} + i\l G\sigma^a\partial_\mu \br G - \partial_\mu G\sigma^a \br G\r~.
\eea
This expression is essentially equivalent to the construction discussed in (\ref{eq_vierbein}) but we will use this notation in this section in order to avoid confusion with the standard literature in the field.\footnote{Notice that in this formalism the goldstino/phonino has different transformation properties. Also, when using the AV formalism we will assume a flat metric.}
This vierbein is used to define the AV covariant derivative
\bea
\AVD_a \phi&\equiv&\AAi^{~\mu}_a\partial_\mu\phi~,
\eea
which transforms according to the standard NLRS. Another well known result is that in order to build a manifestly SUSY invariant Lagrangian out of a non-SUSY Lagrangian, one has to make the following modification
\bea
\label{eq_NLRS_Lagrangian}
\LL(\phi,\partial \phi)
\rightarrow  \AA  \LL(\phi,\AVD \phi).
\eea
In other words, one has to multiply by the determinant of the AV vierbein, and replace all derivatives with the AV covariant derivate.

Thirdly, we would like to discuss another aspect in which the AV formalism enters the low energy theory, even in the absence of a Lagrangian description of the system. As far as we are aware this point does not appear in the literature on the subject.
Theories with symmetries require well defined currents, $j^\mu$, which should satisfy well defined conservation equations. Since the conservation equations contain derivatives and not covariant derivatives, it is not trivial that the SUSY transformation of the operator $\partial_\mu j^\mu$ satisfies the SUSY algebra. In appendix \ref{sec_conservation_NLRS} we show that in order to construct a current for which the SUSY transformation of both the current and its conservation equation satisfies this condition,
one should use the following form
\bea
\label{eq_NLRS_currents}
j^\mu&=&\AA\AAi_a^{~\mu}\hat  j^a
\eea
where $\hat  j^a$ is an operator that in the limit where the goldstino is set to zero is equal to the current, and it is defined to transform according to the standard NLRS (\ref{eq_standard_NLRS}). We will refer to such currents as "standard realization" (SR) currents, bearing in mind that these are merely useful notations and not necessarily physical quantities.
The physical current $j^\mu$ has the same value at the zero goldstino limit, but has a different transformation law.

In appendix \ref{sec_currents} we show that this is the general form for the Noether currents in a theory with NLRS.
However, we would like to stress that it can be used also in theories without a Lagrangian description, because it comes directly from constraints imposed by the SUSY algebra.
More interestingly, in our effective field theory discussion of hydrodynamics we find the same form for the entropy current
\bea
\label{eq_AV_s_current}
s^\mu&=&\AA\AAi_a^{~\mu}\hat  s^a
\eea
and demonstrate that indeed this is a conserved current.

\subsection{A Lagrangian formulation of ideal hydrodynamics with phoninos}

In this subsection we briefly review a Lagrangian formulation of ideal hydrodynamics (see, e.g.,  \cite{Son:2002zn,Dubovsky:2011sj}). We begin with the following Lagrangian
\bea
\label{eq_lag_hydro}
\LL&=&F(B) =F\left(\sqrt {\det (\partial_\mu\phi^I\partial^\mu\phi^J)}\right),
\eea
where we denote $B^{IJ}\equiv \partial_\mu\phi^I\partial^\mu\phi^J$,
$B\equiv\sqrt{\det(B^{IJ})}$, and $F(B)$ is some unspecified function.
$F(B)$ is analogous to $P(T)$ which was introduced in the generating functional formalism, but it has a different physical interpretation which will be presented below.

The $\phi^I$ fields can be viewed as the coordinates for the location of an element of the fluid in a static coordinate system.
$B$ can be written as:
\bea
\label{eq_s_mu}
B& =&\sqrt{-s_\mu s^\mu}\nl
s^\mu&=&\frac{1}{6}\epsilon^{\mu\nu\rho\sigma}\epsilon_{IJK}\partial_\nu \phi^I \partial_\rho\phi^J\partial_\sigma\phi^K~.
\eea
We note that by definition the vector $s^\mu$ has zero divergence.
Also, it is normal to the gradient of the coordinate, so it has to be proportional to the velocity.
Recalling that the velocity is normalized to $u_\mu u^\mu=-1$ we find the relation
\bea
u^\mu&=& \frac{s^\mu}B~.
\eea

The canonical stress energy tensor in this system is
\bea
T^{\mu\nu}
=\eta^{\mu\nu}  F - F_BB\l B ^{-1}\r_{IJ}\partial^\mu\phi^I\partial^\nu\phi^J
&=& \l F-F_BB\r \eta^{\mu\nu}  -F_BBu^\mu u^\nu
\eea
where we used the relation
\bea
\l B ^{-1}\r_{IJ}\partial^\mu\phi^I\partial^\nu\phi^J\equiv
\Delta^{\mu\nu}
=
\eta^{\mu\nu} + u^\mu u^\nu~.
\eea
This relation can be understood by noting that this is a projection operator which satisfies $\Delta^{\mu\nu}u_\mu=0$.
We can identify the isotropic part of the tensor as the pressure and define
\bea
\label{eq_pressure}
P\equiv F-F_BB\qquad\ee\equiv -F \ ,
\eea
where $\ee$ is the energy density and we denote $F_B\equiv \frac{dF}{dB}$.
Using the thermodynamic relations (in the absence of global charges)
\bea
\label{eq_thermo}
Ts=\ee+P \ ,
\eea
one can identify $B$ as the entropy density $s$ and $-F_B$ as the temperature  (up to multiplicative factors).
This means that the current $s^\mu=Bu^\mu$ is in fact the entropy current, and as mentioned above, it is conserved by construction.
This is an important point because it means that the system is not dissipative and therefore the description using a Lagrangian formalism is consistent.

The last missing piece is the differential thermodynamic equation
\bea
\label{eq_thermo_diff}
T d s = d\ee \ ,
\eea
which is necessary to show the consistency of the conservation of the entropy current when the equations of motions are satisfied:
\bea
\label{eq_entropy_conservation}
-T\partial_\mu \l s u^\mu\r=-u^\mu\partial_\mu\ee - \l\ee + P\r\partial_\mu u^\mu
=u_\nu \partial_\mu T^{\mu\nu}=0
\eea

The final form of the action is
\begin{equation}
S=-\int d^4 x\,\varepsilon(s),
\end{equation}
where $s=B$ is the entropy density and $\varepsilon=-F$ is the energy density. Note that this is related to the action we used in the generating functional formalism by a Legendre transform
\begin{align}
\notag &P\left(T=\frac{\partial \varepsilon}{\partial s}\right)=-\varepsilon+s\frac{\partial \varepsilon}{\partial s},\\
&-\varepsilon\left(s=\frac{\partial P}{\partial T}\right)=P-T\frac{\partial P}{\partial T}.
\end{align}
This establishes the connection between the two formalisms presented in this paper. In both cases we define a vector which is proportional to the velocity, $V^\mu$ in the generating functional case and $s^\mu$ in the effective field theory, whose normalization is in fact the basic dynamical variable in the action ($s=\sqrt {-s^2}$ and $T=\frac{T_0}{\sqrt{ -V^2}}$) and it is related via the Legendre transform to the other. Spontaneously broken supersymmetry will enter the game in both cases dressing this vector with modified vierbeins.

The form of the Lagrangian appearing in (\ref{eq_lag_hydro}) was dictated by the symmetries of the system -- translations, rotations and volume conserving diffeomorphisms. Now, we would like to take into account NLRS as well, using the prescription given in (\ref{eq_NLRS_Lagrangian}):
\bea
\label{eq_NLRS_hydro_modification}
F(B)&\rightarrow&\AA F(\BB) \ ,
\eea
where now
\bea
\BB&=&\sqrt{\det (\AVD_a\phi^I\AVD^a\phi^J)}=\sqrt{-\hat s_a \hat s^a}\nl
\hat s^a&=&\epsilon^{abcd}\epsilon_{IJK}\AVD_b \phi^I \AVD_c\phi^J\AVD_d\phi^K
\eea
Using this procedure we find an action which is manifestly invariant under the non-linear supersymmetry transformation (\ref{eq_G_NLRS}) and (\ref{eq_standard_NLRS}), to be compared with the generating functional approach
which is invariant only up to equations of motion under (\ref{transfphonino}).
The advantage of the approach taken in this section is that it contains the phonon and phonino as elementary fluctuations with respect to a static background, and can be used to study their
scattering processes in this limit.

A few comments regarding the vector $\hat s^a$ are in place. First of all, it transforms according to the standard NLRS, because it is constructed using AV covariant derivatives. Secondly, unlike the vector $s^\mu$ defined in (\ref{eq_s_mu}), it is not necessarily conserved (even with respect to an AV covariant derivatives, because the AV covariant derivatives do not commute). Most importantly, it can be related to the vector $s^\mu$ by multiplying by AV factors:
\bea
\label{eq_s_hats_relation}
\AA \AAi_a^{~\mu}\hat s^a
&=&\frac{1}{6}\AA \AAi_a^{~\mu}
\AAi_b^{~\nu}
\AAi_c^{~\rho}
\AAi_d^{~\sigma}
\epsilon^{abcd}\epsilon_{IJK}\partial_\nu \phi^I \partial_\rho\phi^J\partial_\sigma\phi^K=s^\mu\nl
\eea
We will show below that the vector $s^\mu$ can be identified as the entropy current even in the supersymmetric case. This results in two important conclusions: the entropy current in the supersymmetric fluid is conserved by construction,
and its conservation equation satisfies the SUSY algebra. The second point is ensured by the fact that it is written in the form suggested in  (\ref{eq_AV_s_current})\footnote{Notice that the dressing of $\hat s^a$ is similar to the dressing of the vector $V^a$ which was introduced in eq. (\ref{eq_dressing_V}) without the usage of the AV covariant derivative.}.

In order to verify that the vector $s^\mu$ can indeed be identified as the entropy current, we derive the canonical energy-momentum tensor in  (\ref{eq_SE_tensor}), and find that it equals the SR stress-energy tensor dressed with AV factors
\bea
T^{\mu a}&=&\AA\AAi_b^{~\mu} \hat T^{ab}
=\AA \AAi_b^{~\mu}\l  \eta^{ab} \l  F- F_\BB \BB\r - F_\BB \BB \hat u^a \hat u^b\r \ ,
\eea
where we defined $\hat u^a\equiv \frac{\hat s^a}{\BB}$ and identified a projection operator
\bea
\l \BB ^{-1}\r_{IJ}\AVD^a\phi^I\AVD^b\phi^J\equiv
\hat \Delta^{ab}
=
\eta^{\mu\nu} + \hat u^b \hat u^a~.
\eea
Making the identification as in (\ref{eq_pressure}) and (\ref{eq_thermo}) for the hatted objects, we can
define the temperature as $T=-F_\BB$ and the entropy density $ \hat s=\BB$.
In appendix \ref{app_entropy_conservation} we follow a similar computation as in  (\ref{eq_entropy_conservation}) and show that the thermodynamic relations and the conservation of currents indeed lead to the conservation of the entropy current $s^\mu$ also in the supersymmetric case.

As a side remark, we would like to mention a subtlety which is related to the definition of charge densities in this formalism.
Unlike a current, the charge density is a frame dependent quantity. Using the formalism described in this section
one can define two different normalized velocity vectors
\bea
u^\mu\equiv\frac{s^\mu}{B}
\qquad
\hat u^a\equiv\frac{s^a}{\hat B }~.
\eea
$\hat u^a$ transforms according to the standard NLRS, and can be used to define a SR current
\bea
\hat j^a &=& \hat \rho \hat u^a
\eea
assuming that $\hat \rho$ transforms according to the standard NLRS as well. We can also define the physical current using $u^\mu$, but these two definitions have to be related as follows:
\bea
j^\mu=\rho u^\mu &=&\AA \AAi_a^{~\mu}\hat \rho \hat u^a
\eea
Using  (\ref{eq_s_hats_relation}), we find that the two definitions of charge density are related by
\bea
\hat \rho&=& \rho \frac{\BB}{B}~.
\eea

\subsection{Phonino interactions in the fluid}

\label{sec_G_fluctuations}
The Lagrangian formulation of classical ideal hydrodynamics can be used to find the interaction terms of the phonon with the phonino or with itself by expanding around the ground state:
\bea
\phi^I=B_0^{1/3}\l
x^I+\pi^I
\r
\eea
where we identify $\pi^I$ as the phonon, and $B_0$ gives the value of $B$ in the ground state.
The leading order terms in an expansion in the $\pi^I$ fields are:
\bea
\label{eq_lag_phonon}
\delta \LL
&=&\frac{\partial F}{\partial B_{IJ}}\Big|_0\delta B^{IJ}
+\half\frac{\partial^2 F}{\partial B_{IJ}\partial B_{KL}}\Big|_0\delta B^{IJ}\delta B^{KL}
+\ldots\nl
&=&  Ts \l
\half\l \dot \pi^I\r^2-\half c_s^2
\l \partial_I \pi^I \r^2\r +\ldots
\eea
where we used  (\ref{eq_pressure}) and the speed of sound is given by
\bea
c_s^2&=&\frac{\partial P}{\partial \ee}\Bigg|_s=\frac{P'(B)}{ \ee'(B)}=\frac{F_{BB}B}{F_B}~.
\eea
Some details of this computation are given in appendix \ref{app_fluctuations}.
The Lagrangian (\ref{eq_lag_phonon}) can be used to study phonon scattering \cite{Endlich:2010hf} and it is a good starting point to systematically include non-dissipative terms.


We would like to use the same method to discuss the dynamics of the phoninos in an ideal fluid with NLRS, and therefore make the modification appearing in  (\ref{eq_NLRS_hydro_modification}). Taking into account the fluctuations in the AV matrix
\bea
\delta\AA^{~a}_{\mu}
&=&i
\l G\sigma^a\partial_\mu \br G - \partial_\mu G\sigma^a \br G\r~,
\eea
we find new interaction terms in the expansion of the Lagrangian around the static background:
\bea
\label{eq_phonon_goldstino_fluctuations}
\delta \l \AA\LL\r&=&\frac{\partial \l \AA\LL\r}{\partial \AA^{~a}_{\mu}}\Bigg|_0\delta\AA^{~a}_{\mu}+ \frac{\partial^2 \l \AA\LL\r}{\partial\AA^{~a}_{\mu}\partial \BB_{IJ}} \Bigg|_0\delta\AA^{~a}_{\mu}\delta \BB_{IJ}+\ldots
\eea

The first term is the phonino kinetic term
\bea
i\frac{\partial \l \AA\LL\r}{\partial\AA^{~a}_{\mu}}\Bigg|_0
\l G\sigma^a\partial_\mu \br G - \partial_\mu G\sigma^a \br G\r
=i\eta^{b\mu}{\hat T_{ab}}\l G\sigma^a\partial_\mu \br G - \partial_\mu G\sigma^a \br G\r~.
\eea
 where we used (\ref{eq_classical_SE_tensor}).
 This kinetic term leads to the phonino dispersion relations discussed in the introduction. We see also that the canonically normalized phonino should be divided by a factor of $\sqrt {2{\hat T_{00}}}=\sqrt {2\ee}$.

The next term in  (\ref{eq_phonon_goldstino_fluctuations}) gives the two-goldstinos one-phonon vertex:
\bea
\label{eq_phonon_2phonino_vertex}
\delta\l \AA\LL\r&\supset&
 i Ts
\l\l G\sigma^0\dot{ \br G }- \dot{G} \sigma^0\br G\r
+c_s^2\l G\sigma^I\partial_I \br G - \partial_I G\sigma^I \br G\r
\r \l\partial_J \pi_J\r\nl
&&-i Ts\l G\sigma^I\partial_J \br G - \partial_J G\sigma^I \br G\r\l \partial_I\pi_J + \partial_J\pi_I\r \ .
\eea
Details of the computations and computations of interaction terms with a higher number of phonons appear in appendix \ref{app_fluctuations}.
Notice that the phonino does not interact with the transverse mode of the phonon (or equivalently, the vortex $\partial \times \pi$).
Using the canonically normalized phonino, we find that the effective coupling for the phonon-goldstino vertex (and similarly, the scattering of the phonino off $n$ phonons) is proportional to $ Ts/ \ee $. If the temperature is very low in comparison with the scale of supersymmetry breaking $\sqrt{f}$, the coupling will be very small as well, since $\varepsilon \sim |f|^2 \gg T s$. On the other hand, at high temperatures the coupling will be of order one. Therefore, as the temperature is increased the phonino becomes more strongly coupled with the fluid.

\section{Discussion}

We presented the formalism with which the fermionic zero mode associated with the spontaneous breaking of supersymmetry enters the hydrodynamics description of a fluid.
The main ingredient was the dressing of the thermodynamics variables with a vierbein constructed using the phonino/goldstino.
This was demonstrated in two frameworks -- the generating functional approach for hydrodynamics, and an effective field theory with non-linearly realized supersymmetry.

The generating functional approach enabled us to write a hydrodynamics description for the supersymmetric fluid in the presence of external sources. We considered the first non-dissipative derivative corrections. 
The supercurrent first derivative terms take the general form \cite{Kovtun:2003vj}
\begin{equation}
S_i=-D_s\cD_i \bar{G}-D_\sigma\sigma^{ij}\cD_j \bar{G} \ .
\end{equation}
The transport coefficients $D_s$ and $D_\sigma$ correspond to are our $P_2$ and $P_3$ in (\ref{Ps}), which shows that they are non-dissipative. Note that $D_s\propto i P_2$ and $D_\sigma \propto i P_3$, so the coefficients can be complex in general.
 In a conformal theory the two transport coefficients are real and have the same value $D_s=D_\sigma$.

We have limited ourselves to the discussion of the energy-momentum tensor and the supercurrents, a clear extension will be to include conserved currents for global symmetries. In the generating functional formalism this can be implemented by including gauge fields and their fermionic superpartners in the supersymmetry transformations. More generally, the approach can be extended to larger supersymmetries, so that the sources transform as components of multiplets of different gauged supergravity theories.

The effective field theory approach enabled us to write scattering vertices for the phoninos.
Another useful result is a new prescription for dressing conserved currents with goldstinos in effective field theory.
One simply writes the currents using objects which transform according to the standard realization, and multiplies by the appropriate factors of the AV theory:
\bea
\label{eq_conclusions}
T^{\mu a}&=&\AA\AAi_b^{~\mu} \hat T^{ab}\nl
j^{\mu}&=&\AA\AAi_b^{~\mu} \hat j^{b}\nl
s^{\mu}&=&\AA\AAi_b^{~\mu} \hat s^{b} \ .
\eea
This form ensures that the transformation of both the current and its divergence satisfies the SUSY algebra.
The consistency of this prescription with thermodynamics in the ideal case was a non-trivial test for its validity.

Let us now comment on possible phenomenological implications of the results presented in this paper.
The low energy theorems for theories with spontaneously broken symmetry tell us that the interaction between the goldstino and the supercurrent is suppressed by the SUSY breaking scale.
However, in the paradigm of SUSY breaking in a hidden sector, it is well known that the effective interactions with the standard model particles should be suppressed by a higher scale, the mediation scale $M$, to ensure the decoupling of the goldstino in the $M\to\infty$ limit \cite{Leigh:1995jw}. In other words, when taking into account all quantum corrections, there should be some cancelations that cause the coupling of the gravitino to the SM plasma to be suppressed by $\frac{1}{M}$.
However, we suggest that this should be reconsidered. One may view the low energy fermionic zero mode as a linear combination of the hidden sector goldstino and the SM phonino. The hidden sector goldstino coupling is indeed suppressed by the mediation scale, but that of the phonino is not.
Thus, although the contribution of the phonino to the wave function is smaller than that of the hidden sector goldstino, its interactions dominate and lead to the scattering terms discussed in section \ref{sec_G_fluctuations}. Such enhanced interactions with the low energy modes of the plasma should  be taken into account in the study of gravitino physics in the early universe, and perhaps even modify its production mechanism. To our knowledge, this effect was not considered in the literature to date.

Another possible phenomenological consequence of this analysis might be in relaxing the constraints on the gravitino warm dark matter scenario:
It is standard lore that dark matter cannot be a thermal gravitino relic -- such a relic must have mass of the order of 0.1 KeV to explain the dark matter mass density,
while a thermal relic of the same mass would have a velocity distribution which is inconsistent with limits obtained from large scale structure \cite{Kolb:1990vq}.
The lower bound on the mass of such a relic arising from large scale structure is of the order of several KeV, merely one order of magnitude higher than the mass suggested by the relic density. In fact, a borderline scenario of a KeV gravitino as a warm dark matter candidate has recently gained some renewed interest, since it provides a drop in the power spectrum which might be able to solve
astrophysical inconsistencies such as the core vs. cusp question and the emptiness of voids (for a review we suggest \cite{deVega:2010wj}).
One possible solution is to consider a 0.1 KeV gravitino which freezes out with the correct relic density, but with a velocity distribution which is modified by phonino interactions.
In other words, when the gravitino is thermalized, one has to take into account the phonon-phonino effective interactions in the kinetic theory describing the freeze-out process of the gravitino.
Since these interaction is proportional to the momentum of the phonon, this effect might slow down the phonino, and perhaps even lead to the required drop in the power spectrum.
We leave this idea for future work.

\section*{Acknowledgements}
We thank S. Cremonesi and D. Melnikov for collaboration on the early stages of the work, and K. Jensen, E. Kovetz and A. Yarom for useful discussions.
C.~H. was supported in part by the Israel Science Foundation (grant number 1468/06).
B.~K. was supported by the Swiss National Science Foundation (grant number 200021-125237).

\appendix

\section{Non-linearly realized supersymmetry}

\label{app_NLRS}
In this appendix we review some well known results regarding the Akulov-Volkov (AV) formalism for non-linearly realized supersymmetry (NLRS) \cite{Volkov:1972jx}, and give new general expressions for conserved currents in this framework. In this section we will refer to the fermionic zero mode as a goldstino. We will assume a flat metric, use the conventions of \cite{Wess:1992cp} and absorb the supersymmetry breaking scale into the definition of the goldstino. For additional reading we refer the reader to  \cite{Clark:1996aw,Clark:1988es}.

\subsection{The standard non-linear realization of supersymmetry}

In the standard nonlinear realization of supersymmetry, the goldstino transforms as follows:
\bea
\label{eq_G_NLRS}
\delta_\xi G^\alpha &=&
\xi^\alpha  +\VV^\mu_\xi\partial_\mu G^\alpha\nl
\delta_\xi \br G^\dalpha &=&
\br \xi^\dalpha  +\VV^\mu_\xi\partial_\mu\br G^\dalpha~,
\eea
where $G$ is the goldstino and
\bea
\VV ^\mu_\xi&\equiv&-i\l G \sigma^\mu \br \xi-\xi \sigma^\mu \br G\r~.
\eea
Due to its non-linear transformation, the goldstino can be used to realize the SUSY transformation of other fields:
\bea
\label{eq_standard_realization}
\delta_\xi\phi&=&
\VV^\mu_\xi\partial_\mu \phi~.
\eea
Indeed, using the transformation of $\VV^\mu_\xi$
\bea
\delta_\eta \VV^\mu_\xi&=&-i \l\eta \sigma^\mu \br \xi-\xi \sigma^\mu \br \eta\r + \VV^\nu_\eta\partial_\nu \VV^\mu_\xi~,
\eea
one can see that (\ref{eq_standard_realization}) satisfies the supersymmetry algebra
\bea
\label{eq_SUSY_algebra}
\l \delta_\xi\delta_\eta-\delta_\eta\delta_\xi\r\phi
&=&
-2i\l \xi \sigma^\mu \br \eta-\eta \sigma^\mu \br \xi\r \partial_\mu \phi~.
\eea
In this discussion we will assume that there is no linear realization of supersymmetry. In other words, the energy scale is lower than the boson-fermion mass splitting, such that only one ``representative'' of a supermultiplet exists in the low-energy spectrum.

A very useful object in this formalism is the  AV vierbein:
\bea
\AA^{~a}_{\mu}&\equiv&\delta^{a}_{\mu} + i\l G\sigma^a\partial_\mu \br G - \partial_\mu G\sigma^a \br G\r,
\eea
whose variation is
\bea
\delta_\xi\AA^{~a}_{\mu}=\VV^\nu_\xi\partial_\nu\AA^{~a}_{\mu}+\AA^{~a}_{\nu}\partial_\mu\VV^\nu_\xi.
\eea

Its determinant $\AA$ transforms as
\bea
\delta_\xi \AA&=&\partial_\mu\l \VV^\mu_\xi\AA\r~,
\eea
therefore the product of $\AA$ with any objects which transforms according to the standard realization will transform into a total derivative. This is most useful when constructing actions invariant under NLRS.

Gradients do not transform according to the standard realization (\ref{eq_standard_realization}). We therefore use the inverse vierbein, $\AAi^{~\mu}_{a}$ ,which transforms as
\bea
\label{eq_AAi_transformation}
\delta_\xi \AAi^{~\mu}_{a}&=&
\VV^\nu_\xi \partial_\nu \AAi^{~\mu}_{a}
- \AAi^{~\nu}_{a}\partial_\nu  \VV^\mu_\xi  ~,
\eea
and define the AV covariant derivative
\bea
\AVD_a \phi&\equiv&\AAi^{~\mu}_a\partial_\mu\phi
\eea
which does transform according to the standard realization
\bea
\delta_\xi\l \AVD_a\phi\r&=&
\VV^\nu_\xi  \partial_\nu\l \AVD_a \phi\r~.
\eea
The inverse vierbein itself can be written using the covariant derivative
\bea
\label{eq_AAi}
\AAi^{~\mu}_{a}&=&\delta_{a}^{\mu} - i\l G\sigma^\mu\AVD_a \br G - \AVD_a G\sigma^\mu \br G\r~.
\eea
We note that the covariant derivatives are not commuting:
\bea
[\AVD_\mu,\AVD_\nu]\phi
&=&-i\l \AVD_\mu G\sigma^\tau \AVD_\nu \br G -
\AVD_\nu G\sigma^\tau \AVD_\mu \br G\r \AVD_\tau \phi~.
\eea

\subsection{Conservation equations in NLRS}

\label{sec_conservation_NLRS}
As explained in the text, conserved currents are unique objects in the sense that their supersymmetry transformation properties are more constrained -- we require the transformation of both the current and its divergence to satisfy the SUSY algebra. Since the derivative in the conservation equations is not the AV covariant derivative, this is not a trivial condition.
We therefore suggest the following form for the currents:
\bea
\label{eq_AV_current}
j^\mu&=&\AA\AAi_a^{~\mu}\hat  j^a
\eea
where $\hat j^a$ is the current defined such that it coincides with the non-SUSY current in the limit where the goldsitno is set to zero, and it transforms according to the standard realization. We refer to these currents as  "standard realization" (SR) currents.
$j^\mu$ defined in this way coincides with the classical current in the $f\to\infty$ limit as well, but it has different transformation properties:
\bea
\label{eq_current_transformation}
\delta_\xi  j^{\mu}
&=&\partial_\nu \l \VV^\nu_\xi j^{\mu}-\VV^\mu_\xi j^{\nu }\r
+\VV^\mu_\xi\partial_\nu j^{\nu}\nl
\delta_\xi \l \partial_\mu j^{\mu}\r &=&
\partial_\mu\l  \VV^\mu_\xi\partial_\nu j^{\nu} \r~.
\eea
Applying another NLRS transformation we find that the $\VV$ dependent terms are symmetric under the exchange of $\xi$ and $\eta$ and will therefore vanish from the commutator:
\bea
\delta_\eta\delta_\xi j^\mu
&=&-i \l\eta \sigma^\nu \br \xi-\xi \sigma^\nu \br \eta\r\partial_\nu j^\mu +\partial_\rho\partial_\nu
\l\VV^\rho_\eta  \VV^\nu_\xi j^{\mu}\r
-\partial_\nu \l \VV^\nu_\xi \partial_\rho  \VV^\mu_\eta j^\rho \r
- \partial_\nu \l \VV^\nu_\eta \partial_\rho \VV^\mu_\xi j^{\rho}\r
\nl
\delta_\eta\delta_\xi \l \partial_\mu j^{\mu}\r
&=&-i\l \xi \sigma^\mu \br \eta -\eta\sigma^\mu\br\xi\r \partial_\mu \l \partial_\nu j^{\nu} \r
+\partial_\mu\partial_\rho \l \VV_\eta^\rho \VV^\mu_\xi\l \partial_\nu j^{\nu} \r\r~.
\eea
The important point is that with this construction the SUSY algebra is indeed satisfied, as in (\ref{eq_SUSY_algebra}), for both the current and its divergence.

In the next section of this appendix we will show that this is the form of the currents which is obtained from the Noether procedure.

\subsection{Invariant actions and Noether currents in NLRS}
\label{sec_currents}

In this section we use the well-known construction of invariant actions in the AV formalism, compute Noether currents, and show that indeed they take the form suggested in   (\ref{eq_AV_current}).
To make a Lagrangian invariant under non-linearly realized SUSY we make the following modification
\bea
\LL(\phi^I,\partial \phi^I)
\rightarrow \AA \LL(\phi^I,\AVD \phi^I)
\eea
(the index $I$ runs over all the fields in the theory)~.
In other words, we replace all derivatives with covariant derivatives and multiply the Lagrangian by $\AA$ to make it transform into a total derivative as explained above .

Let us first consider a Noether current corresponding to a global symmetry.
We can define a ``standard realization'' (SR) current applying the Noether formula to the Lagrangian:
\bea
\hat j^a&\equiv&
 \frac{\partial  \l \LL\l \phi,\AVD\phi\r\r}{\partial (\AVD_a \phi^I)}
\frac{\partial \l \delta_\alpha \phi ^I \r}{\partial \alpha}
\eea
This is, of course, just a useful notation. The physical current is in fact given by
\bea
j^\mu&=&
\frac{\partial  \l \AA \LL(\phi,\AVD \phi)\r}{\partial (\partial_\mu \phi)}
\frac{\partial \l \delta_\alpha \phi  \r}{\partial \alpha}
=\AA\AAi_a^{~\mu}
\frac{\partial  \l  \LL(\phi,\AVD \phi)\r}{\partial (\AVD_a \phi^I)}
\frac{\partial \l \delta_\alpha \phi ^I \r}{\partial \alpha}
=\AA\AAi_a^{~\mu} \hat j^a\nl
\eea
in accordance with the prescription given in the previous section.

We will now compute the stress-energy tensor. For this purpose we define an SR stress-energy tensor:
\bea
\label{eq_classical_SE_tensor}
\hat T^{ab}&\equiv&-\frac{\AAi^{a\mu}}{\AA}\frac{\partial{\l\AA\LL\r}}{\partial \l\AAi_{b}^{~\mu}\r}
=\eta^{ab}\LL
-\frac{\partial{\LL}}{\partial \l \nabla_b \phi^I\r }
\nabla^a\phi^I
\eea
In the limit of zero goldstino fluctuations it is identical to the canonical stress-energy tensor, and it transforms according to standard NLRS because it is constructed from Lorentz scalars and covariant derivatives:
\bea
\label{eq_classical_T_transformation}
\delta_\xi\hat  T^{ab}&=&
\VV_\xi^\mu\partial_\mu \hat T^{ab}~.
\eea

The canonical stress-energy tensor in the presence of goldstinos can be related to the classical one as follows:
\bea
\label{eq_SE_tensor}
 T^{\mu a}&=&
-\partial^a G^\alpha\frac{\partial \l \AA \LL\r}{\partial \l\partial_\mu G^\alpha\r}
-\frac{\partial \l \AA \LL\r}{\partial \l\partial_\mu \br G^\dalpha\r}\partial^a \br G^\dalpha
-\frac{\partial\l \AA\LL\r}{\partial \l\partial_\mu\phi^I\r}\partial^a \phi^I
+\eta^{\mu a}\AA \LL\nl
&=&
-\AAi_b^{~\mu}
\AAi^{a\rho}
\frac{\partial{\l \AA \LL\r}}{\partial \l\AAi_{b}^{~\rho}\r}
\nl
&=&\AA\AAi_{b}^{~\mu} \hat T^{ab}
\eea
where we found the following relation useful:
\bea
-\frac{\AAi_b^{~\mu}}{\AA}\frac{\partial{\l\AA\LL\r}}{\partial \l\AAi_{ba}\r}
&=&\eta^{\mu a}\LL -
\frac{\partial{\LL}}{\partial \l \partial_\mu \phi^I\r }
\partial^a \phi^I~.
\eea
Again, we see that the form of the stress-energy tensor derived from the invariant action agrees with the prescription suggested above, therefore the transformation of both the tensor and its divergence satisfies the SUSY algebra. As in  (\ref{eq_current_transformation}), the NLRS transformation is given by
\bea
\delta_\xi T^{\mu a}
&=&\partial_\nu \l \VV^\nu_\xi T^{\mu a}-\VV^\mu_\xi T^{\nu a}\r
+\VV^\mu_\xi\partial_\nu T^{\nu a}
\eea
This tensor is not symmetric, but it can be made symmetric (on-shell) by adding a Belinfante term
\bea
\Theta^{\mu\nu}&=&T^{\mu\nu}+\half\partial_\rho\l H^{\rho\mu\nu} + H^{\mu\nu\rho}+H^{\nu\mu\rho} \r\nl
H^{\rho\mu\nu}&\equiv&\epsilon^{\mu\nu ab} T^{\rho}_{~ a}G\sigma_b  \br G~.
\eea

We move on to compute the supercurrent using the Noether procedure (as mentioned above, we assume that all fields transform according to the NLRS):
 \bea
 \label{eq_supercurrent}
 S^\mu_{\alpha}&=&
\frac{\partial \l \delta_\xi G^\beta\r}{\partial \xi^\alpha}
\frac{\partial  \l \AA \LL\r}{\partial (\partial_\mu G^\beta)}
+
\frac{\partial \l \AA \LL\r}{\partial (\partial_\mu \br G^\dbeta)}
\frac{\partial ( \delta_\xi \br G^\dbeta)}{\partial \xi^\alpha}
+
\frac{\partial  \l \AA \LL\r}{\partial (\partial_\mu \phi^I)}
\frac{\partial \l \delta_\xi \phi^I\r}{\partial \xi^\alpha}
- i \AA \LL\l\sigma^\mu \br G\r_\alpha
\nl
&=&-2i T^{\mu a }\l\sigma_a \br G\r_\alpha
\eea
Using the transformation properties of the stress-energy tensor and the goldstino we find
\bea
\delta_{\xi} S^\mu_\alpha &=&
\partial_\nu \l \VV^\nu_\xi S^{\mu}_\alpha - \VV^\mu_\xi S^{\nu }_\alpha\r
+\VV^\mu_\xi\partial_\nu S^{\nu}_\alpha
-2i T^{\mu \nu}\l \sigma_\nu \br\xi\r_\alpha~.
\eea
As required by the SUSY algebra, the supercurrent indeed transforms into the stress-energy tensor (plus derivative terms). Substituting (\ref{eq_SE_tensor}) into (\ref{eq_supercurrent}) we see that the supercurrent can also be written in terms of SR quantities dressed with appropriate goldstino factors.

The conservation equations for the stress-energy tensor and the supercurrent combine to give the equation of motion for the goldstino:
\bea
\label{eq_goldstino_eom}
\AA\AAi_{b}^{~\mu} \hat T^{ab}\partial_\mu\l \sigma_a  \br G\r_\alpha = 0
\eea
which reproduces the result appearing in  (\ref{susycond}).

For completeness we also write the expression for the $R$-current in cases in which only the goldstinos have $R$ charge (e.g. hydrodynamics with NLRS, as discussed below):
\bea
\label{eq_R_current}
j_R^\mu &=&-2 T^{\mu a }G \sigma_a \br G\nl
\delta_\xi j_R^\mu
 &=&
\partial_\nu \l \VV^\nu_\xi  j_R^\mu - \VV^\mu_\xi  j_R^\nu\r
+\VV^\mu_\xi\partial_\nu  j_R^\nu
-i \xi S^\mu +i\br \xi\br S^\mu ~.
\eea

The results of this section are consistent with \cite{Clark:1988es} when using the goldstino self interactions Lagrangian $\AA\LL=-f^2\AA$.

\section{Conservation of entropy current in fluids with phoninos}
\label{app_entropy_conservation}

As a consistency check for the identification of the entropy current in the supersymmetric case with the vector $s^\mu$ defined in \eqref{eq_s_mu} , we show that the entropy is conserved by virtue of the thermodynamic relations and the equations of motion. We begin by writing the divergence of the entropy current as in (\ref{eq_entropy_conservation}) using the definition $s^\mu=\AA \AAi_a^{~\mu} \hat u^a \hat s$ and the thermodynamic relations (\ref{eq_thermo}) and (\ref{eq_thermo_diff}) for the SR objects:
\bea
-T\partial_\mu s^\mu &=&- \partial_\mu \l \AA \AAi_{a}^{~\mu} \hat u^a\r T \hat s -  \AA \AAi_{a}^{~\mu} \hat u^aT\partial_\mu \hat s\nl
&=&-\partial_\mu \l \AA \AAi_a^{~\mu} \r\l \ee+P\r \hat u^{ a}-\AA\AAi_a^{~\mu}\l \ee+ P\r \partial_\mu \hat u^a
- \AA \AAi_a^{~\mu}\hat u^a \partial_\mu  \ee\nl
&=&\hat u_a \partial_\mu\l \AA \AAi_b^{~\mu}\hat  T^{ab}\r
-\hat u^{ a}\partial_\mu \l \AA \AAi_a^{~\mu} \r P
~.
\eea
Using the relation
\bea
\label{eq_d_A_AAi}
\partial_\mu\l \AA\AAi_a^{~\mu} \r
  &=&2i\AA  \eta^{bc} \l \AVD_a G \sigma_c \AVD_b \br G- \AVD_b G \sigma_c  \AVD_a \br G \r
\eea
we find
\bea
-T\partial_\mu \l  s u^\mu\r
&=&\hat u_a \partial_\mu\l \AA \AAi_b^{~\mu}\hat  T^{ab}\r
-2 i \hat u^a\AA\hat T^{bc} \l\AVD_a G \sigma_c \AVD_b\br G-\AVD_bG\sigma_c\AVD_a \br G\r
\eea
(Notice that the expression proportional to $\hat u^a \hat u^b \hat u^c$ drops from this expression because the RHS in (\ref{eq_d_A_AAi}) is antisymmetric under exchange of $a$ and $b$).
Denoting $
\hat \mu\equiv \hat u^a \AVD _a G
$ \footnote{This notation is based on (\ref{eq_hat_mu}).}, and
recalling  (\ref{eq_SE_tensor}) and (\ref{eq_supercurrent})
we can rewrite this expression as:
\bea
-T\partial_\mu \l  s u^\mu\r
&=&\hat u_a \partial_\mu T^{\mu a}
-2 i  T^{\mu a} \l\hat\mu \sigma_a \partial_\mu\br G- \partial_\mu G\sigma_a\bar {\hat \mu}\r\nl
&=&
\l \hat u_a+2i\l{\hat \mu} \sigma_a \br G-G\sigma_a\br {\hat \mu}\r\r \partial_\mu T^{\mu a}+{\hat \mu} \partial_\mu S^\mu+\br {\hat \mu} \partial_\mu \br S^\mu=0
\eea

\section{Phonon and phonino fluctuations}
\label{app_fluctuations}

In this appendix we collect some results we found useful for the computations of phonon and phonino interactions.

The Lagrangian (\ref{eq_lag_phonon}) by considering the following fluctuation in the matrix $B^{IJ}\equiv \partial_\mu\phi^I\partial^\mu\phi^J$
\bea
\delta B_{IJ}&=&B_0^{2/3}\l \partial_I \pi_J +\partial_J \pi_I+\partial_\mu \pi_I \partial^\mu \pi_J\r
\eea
The necessary derivatives of $F$ are
\bea
\frac{\partial F}{\partial B^{IJ}}\Big|_0&=&\half F_B B \BBi_{IJ}\Big|_0=\half F_B B_0^{1/3} \delta_{IJ} \nl
\frac{\partial^2 F}{\partial B^{KL}\partial B^{IJ}}\Big|_0
&=&\frac 1 4 \l F_{BB}B^2+F_B B \r \BBi_{IJ}\BBi_{KL}
-\half F_B B \BBi _{IK}\BBi_{LJ}\Big|_0\nl
&=&
\frac 1 4 \l F_{BB}B_0^2+F_B B_0 \r B_0^{-4/3}\delta_{IJ}\delta_{KL}
-\half F_B B_0^{-1/3}\delta _{IK}\delta_{LJ}
\eea
Upon integration by parts, a term $\partial_I\pi_J\partial_J \pi_I$ becomes $\l\partial_I\pi_I\r^2$, giving the action \eqref{eq_lag_phonon}.

For the computation of the two-goldstinos one-phonon vertex, (\ref{eq_phonon_2phonino_vertex}), we use the following results:
\bea
\frac{\partial \BB_{IJ}}{\partial \l\AA_{\mu}^{~a}\r}\Bigg|_0
&=&
-\BB_0^{2/3}\l \eta_{a I}\delta^\mu_J+\eta_{a J}\delta^\mu_I\r \nl
\frac{\partial \BB}{\partial \l\AA_{\mu}^{~a}\r}\Bigg|_0&=&
-
\BB_0\delta_a^\mu \l 1-\delta_a^0\r\nl
\frac{\partial^2 \l \AA\LL\r}{\partial\AA^{~a}_{\mu}\partial \BB^{IJ}} \Bigg|_0&=&
\half F_B \BB_0\BB_0^{-2/3} \l \delta_a^\mu \delta_a^0\delta_{IJ}
-c_s^2  \delta_{IJ}\delta_a^\mu \l 1-\delta_a^0\r
+
\l \eta_{a I}\delta^\mu_J+\eta_{a J}\delta^\mu_I\r\r \nl
\eea

Computations of interaction terms with more phonons have an additional subtlety as they involve contributions from different orders in $\delta B_{IJ}$. For example, the goldstino interactions with two phonons has a contribution from the linear order in $\delta B_{IJ}$
\bea
\delta\l
\AA\LL\r&\supset&
\frac{i }{2}Ts \l
\l G\sigma_0\dot{ \br G }- \dot{G} \sigma_0\br G\r
+c_s^2\l G\sigma^I\partial_I \br G - \partial_I G\sigma^I \br G\r
\r \l\partial_\mu \pi _J \partial^\mu \pi_J\r\nl
&&-i Ts \l G\sigma^I\partial_J \br G - \partial_J G\sigma^I \br G\r\partial_\mu \pi _I \partial^\mu \pi_J~,
\eea
and another term from the $\delta B_{IJ}\delta B_{KL}$ order
\bea
\AA\LL&\supset&
\frac{i}{2}Ts \l
\l1+c_s^2\r\l G\sigma_0\dot{ \br G }- \dot{G} \sigma_0\br G\r
+d_s^2\l G\sigma^I\partial_I \br G - \partial_I G\sigma^I \br G\r
\r \l\partial_J \pi_J\r^2\nl
&&+\frac{i}{2}Ts \l
 \l G\sigma_0\dot{ \br G }- \dot{G} \sigma_0\br G\r
+ c_s^2\l G\sigma^I\partial_I \br G - \partial_I G\sigma^I \br G\r
\r \l\partial_J \pi_K\partial_J \pi_K+\partial_J \pi_K\partial_K \pi_J\r\nl
&&-iTs
  \l 1+c_s^2\r \l G\sigma^I\partial_J \br G - \partial_J G\sigma^I \br G\r\l \partial_I \pi_J +\partial_J \pi_I\r\l \partial_K\pi_K\r\nl
&&+\frac{i}{4}Ts \l G\sigma^I\partial_J \br G - \partial_J G\sigma^I \br G\r\l \partial_K \pi_I +\partial_I \pi_K\r\l \partial_K \pi_J +\partial_J \pi_K\r
\eea
where we defined $d_s^2 \equiv \l F_{BBB}B^2+2F_{BB}B\r /F_B$ and used
\bea
 \frac{\partial^3 \l \AA\LL\r}{\partial\AA^{~a}_{\mu}\partial \hat B^{KL}\partial\hat B^{IJ}}\Bigg|_0
 &=&\frac 1 4\BB_0^{-4/3}\delta_a^\mu\delta_a^0
 \l  \l F_{BB}B^2+F_B \hat B \r \delta_{IJ}\delta_{KL}
-\half F_B \hat B \delta _{IK}\delta_{LJ}\r
 \nl
&&- \frac 1 4\hat B_0^{-4/3}\delta_a^\mu \l 1-\delta_a^0\r\l \l F_{BBB}\hat B^3 + 2F_{BB}\hat B^2\r \delta_{IJ}\delta_{KL}
 -\frac{1}{2}  F_{BB} \hat B^2\delta_{IK}\delta_{LJ}\r\nl
 &&
 + B_0^{-4/3}\l F_{BB}\hat B^2+F_B \hat B \r
\eta_{a I}\delta^\mu_J\delta_{KL}
+\frac 1 2\hat B_0^{-4/3} F_B \hat B
\eta_{a I}\delta^\mu_K\delta_{LJ}
\eea

   \end{document}